%% file: paper.tex
\documentclass[10pt,twocolumn,letterpaper]{article}
\usepackage[usenames, dvipsnames]{xcolor}
\usepackage{usenix2019, times,color,graphicx,amsmath,amsthm,array,amssymb, subcaption}
\usepackage{mathptm,times,helvet,courier,xspace}
\usepackage{boxedminipage,multirow,endnotes,balance}
\usepackage{pdfpages}
\usepackage{rotating}
\usepackage{colortbl}
\usepackage{transparent}
\usepackage{appendix}
\usepackage{multirow}
\usepackage{relsize}
\usepackage{booktabs}
\usepackage{multirow}
\usepackage{cite}
\usepackage[font={small,bf}, skip=1pt]{caption}
\usepackage{cancel}
\usepackage{xcolor}
\usepackage[linesnumbered,ruled,vlined]{algorithm2e}
\usepackage{soul}
\usepackage{amsmath}
\usepackage{paralist}
\usepackage{enumitem}
\usepackage{tikz}

\usetikzlibrary{arrows.meta,positioning,shapes,fit}

\DeclareRobustCommand\numcircledtikz[1]{\tikz[baseline=(char.base)]{
    \node[shape=circle,draw,fill,inner sep=1pt] (char)
    {\textcolor{white}{#1}};}}

\SetCommentSty{mycommfont}
\SetKwInput{KwInput}{Input}                
\SetKwInput{KwOutput}{Output}              

\usepackage{listings}

\usepackage[]{hyperref}
\usepackage{xurl}

\newcommand{\nop}[1]{}

\newcommand{\tightcaption}[1]{\vspace{-5pt}\caption{{\bf \small #1}}
\vspace{-15pt}
}
\newcommand{\para}[1]{\smallskip\noindent{\bf #1}}

\newcommand{\cut}[1]{}

\newcommand{\name}{Aragog\xspace}

\def\compactify{\itemsep=0pt \topsep=0pt \partopsep=0pt \parsep=0pt}
\let\latexusecounter=\usecounter

\theoremstyle{definition}

\usepackage{ifthen}
\usepackage{fancyhdr}

\newfont{\dft}{phvb at 6pt}
\newfont{\mft}{phvro at 6pt}
\newfont{\df}{phvb at 9pt}
\newfont{\mf}{phvro at 9pt}
\usepackage{tikz}

  {\begin{list}{$\bullet$}%
     {\setlength{\parsep}{0pt}%
      \setlength{\topsep}{0pt}%
      \setlength{\itemsep}{0pt}}}%
  {\end{list}}
  
\makeatletter

\global\def\section{\@startsection {section}{1}{\z@}%
                                   {-1.5ex \@plus -0.8ex \@minus -.1ex}%
                                   {0.6ex \@plus.2ex}
                                   {\normalfont\bfseries\fontsize{11}{13}\selectfont}}

\global\def\subsection{\@startsection{subsection}{2}{\z@}%
                                     {-1.25ex\@plus -0.8ex \@minus -.1ex}%
                                     {0.3ex \@plus .1ex}
                                     {\normalfont\bfseries\fontsize{10}{12}\selectfont}}
\global\def\subsubsection{\@startsection{subsubsection}{3}{\z@}%
                                     {-1ex\@plus -1ex \@minus -.1ex}%
                                     {0.1ex \@plus .1ex}
                                     {\normalfont\itshape\fontsize{10}{12}\selectfont}}

\usepackage{bibspacing}
\def\noeditingmarks{}
\newcommand{\textred}[1]{\textcolor{red}{#1}}
\ifx\noeditingmarks\undefined
   \newcommand{\pgwrapper}[2]{\textred{#1: #2}}
\else
   \newcommand{\pgwrapper}[2]{}
\fi

\usepackage{epsfig,comment}
\newcommand{\squishlist}{
   \begin{list}{$\bullet$}
    { \setlength{\itemsep}{0pt}      \setlength{\parsep}{3pt}
      \setlength{\topsep}{3pt}       \setlength{\partopsep}{0pt}
      \setlength{\leftmargin}{1.0em} \setlength{\labelwidth}{1em}
      \setlength{\labelsep}{0.5em} } }
\newcommand{\squishend}{
    \end{list}  }

\date{}
\begin{document}

\twocolumn[\begin{@twocolumnfalse}

\begin{centering}

\begin{centering}

{\Large \bf \name{}: Just-in-Time Model Routing for Scalable Serving of Agentic Workflows}\\\vspace{0.25cm}

\def\refPrinceton{$^\mathsection$}
\def\refGatech{$^\dagger$}

{\normalsize 
Yinwei Dai\refPrinceton{}\hspace{0.5cm}%
Zhuofu Chen\refPrinceton{}\hspace{0.5cm}%
Anand Iyer\refGatech{}\hspace{0.5cm}%
Ravi Netravali\refPrinceton{}\\}
\vspace{0.1cm}
{\normalsize \refPrinceton{}Princeton University\hspace{0.5cm}%
\refGatech{}Georgia Institute of Technology\\}
\vspace{0.1cm}

\end{centering}

\end{centering}

\vspace{\baselineskip}

\end{@twocolumnfalse}]

\interfootnotelinepenalty 100000
\widowpenalty 100000
\clubpenalty 100000
\newfont{\tf}{phvro at 9.5pt}
\newfont{\tft}{phvro at 7.25pt}
\begin{sloppypar}

\input{abs}
\input{intro2}
\input{background_v1}
\input{design_v2}

\input{implementation}
\input{eval_v1}

\input{related}
\input{conclusion}
\label{lastpage}
\balance
\Urlmuskip=0mu plus 1mu\relax
\bibliographystyle{abbrv}
\bibliography{paper}

\end{sloppypar}
\clearpage
\sloppypar
\end{document}

%% file: abs.tex
\begin{abstract}
\noindent
Agentic workflows have emerged as a powerful paradigm for solving complex, multi-stage tasks, but serving them at scale is computationally expensive given the many LLM inferences that each request must pass through. Configuration selection, or the cost-aware assignment of workflow agents to specific LLMs, can reduce these costs, but existing approaches bind configuration decisions before request execution, making them ill-suited for the heterogeneous and lengthy execution of workflows. Specifically, system loads can fluctuate rapidly and substantially during a request's lifetime, causing fixed configurations to quickly become suboptimal. We present \name{}, a system that progressively adapts a request's configuration throughout its execution to match runtime dynamics. To make this practical despite the massive space of workflow configurations, \name{} decouples the problem into two core elements -- a one-time routing step that identifies all accuracy-preserving configurations, and a cheap per-stage scheduler that selects among them using up-to-date system observations -- and introduces novel strategies to accelerate each. Across diverse workflows and model families, \name{} increases maximum serving throughput by 42.8--217.0\% and reduces median latency by 32.5--86.1\% at peak request rates, while maintaining accuracy comparable to the most expensive configurations.

\end{abstract}

%% file: intro2.tex
\section{Introduction}
\label{s:intro}

Autonomous agents, or Large Language Models (LLMs) augmented with tools, memory, and planning, have shown impressive ability to tackle complex tasks autonomously~\cite{yang2024sweagent,snowflake-agentic-workflow}. However, fully autonomous agentic systems often lack the structure and reliability needed in production~\cite{multiagentllmsystemsfail}. This has led to the rise of agentic workflows: directed acyclic graphs with fixed structures that orchestrate multiple agents~\cite{langchain,dspy,autogen,metagpt}, each responsible for a well-defined subtask and passing outputs downstream (Figure~\ref{fig:workflow_examples}). These workflows preserve the generative strengths of agents while offering the observability and debuggability of traditional software. Yet serving them at scale remains expensive, largely due to the inflated number of LLM invocations per request. Indeed, a single workflow execution typically triggers several agent invocations, each making multiple LLM calls; the net effect is that LLM inference can account for nearly 80\% of end-to-end request time in workflows (\S\ref{ss:workflows}).

The predominant strategy for reducing these inference costs focuses on intelligently selecting \emph{workflow configurations} -- i.e., assignments of specific LLMs to each agent in the workflow -- that minimize cost while preserving accuracy. Existing systems make configuration decisions either periodically offline~\cite{cognify,llmselector,murakkab} or once per input at request arrival time~\cite{masrouter,routellm,daao}. Regardless, they all commit to configurations before executing a request, focusing on optimizing fixed cost metrics such as API dollars or FLOPs. Unfortunately, such a priori decision making is fundamentally misaligned with the heterogeneous and lengthy nature of workflows: as a request slowly moves through a multi-stage graph, the overall system load it encounters can rapidly fluctuate due to the diverse overheads each agent imposes, and the arrival/departure of other requests. Consequently, a configuration selected at the start may quickly become suboptimal during a request's lifecycle, e.g., an agent assigned a small model to save FLOPs may perform worse than necessary (latency- and throughput-wise) if that model is overloaded and a larger model is idle when the request reaches the agent.

To make matters worse, this early-binding behavior fails to fully exploit the \emph{enhanced configuration flexibility} that workflows naturally afford through multi-stage processing. Specifically, we observe that workflows exhibit substantial fault tolerance -- errors introduced at intermediate stages can often be corrected downstream, e.g., via refinement agents. This, in turn, dramatically expands the set of configurations that preserve end-to-end accuracy, providing a means to cope with (and support) the diverse runtime scenarios that could be encountered during serving. Yet, by binding configurations early, existing approaches forego such opportunities, leaving 25-70\% of the potential gains from configuration adaptation on the table for diverse agentic workloads (\S\ref{ss:limitations}).

We present \textbf{\name{}}, an agentic workflow serving system that capitalizes on the aforementioned flexibility by performing stage-wise configuration adaptation.\footnote{In this paper, we use the terms stage and agent interchangeably to refer to individual execution steps in an agentic workflow.} The primary challenge is one of runtime overheads: per-stage adaptation is prohibitively expensive because it requires frequently exploring an exponentially large configuration space that scales with workflow size and model options. To handle this, our key insight is that, for a given input, the accuracy of a configuration is fixed, whereas its runtime performance can change dramatically as system load fluctuates. \name{} leverages this asymmetry by decomposing configuration selection into two complementary steps. It first performs the expensive, accuracy-related analysis once per request (i.e., when it arrives) to identify all configurations that preserve accuracy. Then, as the request moves through the workflow, \name{} rapidly selects among only the pre-validated configurations for each upcoming stage, using current runtime signals to stay aligned with system conditions.

To realize this decoupling approach while maximizing serving throughput and avoiding any added latency from configuration adaptation, \name{} embeds two optimized components. First, to identify the set of accuracy-preserving configurations, \name{} decomposes the routing problem into many simple binary classification problems and exploits the near-monotonic relationship between configuration FLOPs and accuracy to prune configuration space quickly; this keeps routing both accurate and lightweight enough to overlap with request queuing delays. Second, \name{}'s (stage-wise) runtime scheduler leverages how FIFO priorities and pruned configuration sets naturally structure the assignment space: earlier requests in the queue have only a few viable configurations -- since each choice at an earlier stage can invalidate options that use a different model for that stage -- and once chosen, they constrain later requests rather than letting the space fan out. This, in turn, creates a compact search space that beam search can traverse efficiently, avoiding both exhaustive exploration and greedy myopia.

We evaluate \name{} by comparing with existing configuration optimizers -- both per-workflow~\cite{cognify,llmselector,murakkab} and per-input~\cite{routellm,masrouter,daao} -- augmented with oracle accuracy information about configurations, i.e., representing their best possible performance. Across four popular agentic workflows and diverse model families (Qwen 2.5~\cite{qwen}, Llama 3~\cite{llama3}, and Phi 4~\cite{phi4}), we find that \name{} improves maximum serving throughput by 42.8--76.3\% over per-input optimizations and 78.1--217.0\% over per-workflow optimizations. Further, \name{} consistently reduces latency across varying request rates. For instance, at peak load, median and P95 reductions were 32.5--71.1\% and 46.2--76.2\% over per-input optimizations, and 60.0--86.1\% and 63.2--89.0\% over per-workflow optimizations. Crucially, \name{} achieves these performance wins while maintaining accuracy within 2\% of that when always using the most expensive configuration. We will open source \name{} post publication.

%% file: background_v1.tex
\section{Background and Motivation}
\label{s:background}

We begin by providing an overview of agentic workflows (\S\ref{ss:workflows}), highlighting the limitations of current optimization approaches and opportunities they overlook (\S\ref{ss:limitations}). We then discuss the practical challenges in realizing them (\S\ref{ss:challenges}).

\subsection{Agentic Workflows}
\label{ss:workflows} 

\begin{figure}[t]
    \centering
    \includegraphics[width=\linewidth]{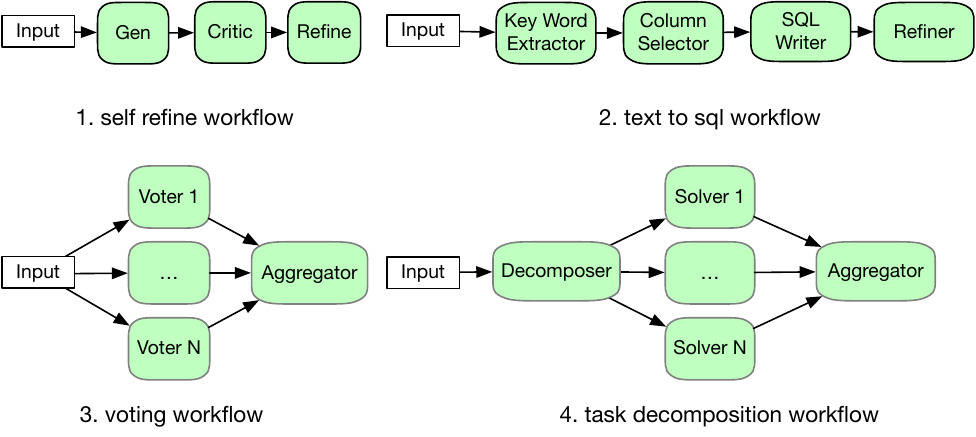}
    \caption{Examples of common agentic workflows that vary in terms of both structure and application domains.}
    \label{fig:workflow_examples}
     \vspace{-10pt}
\end{figure}

Agentic workflows are directed graphs of agent invocations, where each agent---a prompted LLM augmented with tools, memory, and planning capabilities---executes a well-scoped sub-task and passes its output to downstream agents. As shown in Figure~\ref{fig:workflow_examples}, example patterns and application domains include: (1) self-refine workflows~\cite{self_refine,llmselector} that iteratively improve outputs through generation, critique, and refinement agents; (2) natural language to SQL systems~\cite{chess,hexgentext2sql} that decompose query construction into keyword extraction, column selection, SQL generation, and refinement; (3) task decomposition workflows~\cite{decompose,autogen} that split complex problems into parallel sub-tasks handled by different agents before aggregating their results; and (4) multi-agent voting systems~\cite{self-consistency,voting} that ensemble multiple agents for robust decision-making. In practice, these workflows are explicitly encoded using frameworks like LangChain~\cite{langchain} and DSPy~\cite{dspy}. Once a workflow graph is defined, developers can specify a \textit{configuration} that assigns each agent to a chosen LLM, enabling accuracy-cost tradeoffs without modifying workflow logic. These configurations often utilize heterogeneous models -- i.e., smaller models for easier agent tasks, specialized models for domain-specific tasks  -- that are each hosted on different serving engines and GPUs~\cite{llmselector}.

Yet despite these intuitive benefits, agentic workflows face significant practical challenges, largely centered around the high costs they bring. Indeed, each agent invocation incurs the inference cost of powerful yet resource-intensive LLM calls. These expenses scale quickly since workflows generally involve multiple sequential or parallel agent calls. For instance, coding workflows require 3–5 agent calls per query to plan, generate, test, and verify solutions~\cite{zhang2023agents,yang2024sweagent}, while even simple workflows such as multi-hop question answering for legal or scientific search can involve 4+ agent calls~\cite{lewis2020rag,bai2023constitutional,orca}. Across our workloads, a request invokes 5.2 LLM calls on average. These LLM calls account for 79.4\% of end-to-end latency on average, with tool calls and framework (e.g., DSPy) overhead accounting for the remaining 20.6\%.
 
\subsection{Limitations of Existing Optimizations}
\label{ss:limitations}

Existing systems primarily focus on configuration selection to reduce these LLM inference costs while preserving workflow accuracy. We categorize and discuss these techniques below, highlighting their limitations; complementary optimizations are covered in \S\ref{s:related}.

\para{Per-workflow optimization.} Systems like Cognify~\cite{cognify}, LLMSelector~\cite{llmselector}, and Murakkab~\cite{murakkab} perform one-time or periodic configuration selection based on sample data, minimizing costs (e.g., FLOPs, API costs) while preserving accuracy relative to the most expensive configuration. Once selected, a configuration is fixed for all requests, eliminating runtime overhead but sacrificing per-input adaptability.

\para{Per-input optimization.} In single-LLM serving, existing LLM \emph{routing} approaches~\cite{routellm, mixllm} leverage input heterogeneity by training routers to forward each input to the most cost-efficient model that can accurately respond to that input. Recent works extend this routing paradigm to agentic workflows~\cite{daao, masrouter} by selecting not just singular models, but a workflow configuration for each input. Routers are often transformer-based and are trained using accuracy labels relative to the most expensive configuration, enabling
cost reduction while preserving accuracy.

\para{The problem.}
Though effective, existing optimizations fail to adapt to the intrinsic runtime dynamics of workflows, falling short in two key ways. First, they \emph{bind configurations before workflow execution}, i.e., either once before all inputs, or at the start of each input. The core issue is that as a request is executing, system states can change rapidly, rendering these a priori selections suboptimal by execution time. Natural LLM dynamics already introduce significant variance: output lengths vary unpredictably, queue depths fluctuate continuously, and dynamic batching reshapes execution efficiency moment by moment~\cite{orca}. The lengthy, multi-stage nature of agentic workflows only exacerbates this; the load for each concurrent request can change dramatically as it shifts between workflow agents with different overheads (and potentially different models), and the set of concurrent requests can fluctuate many times as a request slowly passes through the entire workflow graph. Consequently, upfront configuration selections can increasingly become suboptimal as requests progress through multi-stage execution.

Second, these same dynamics make the approach of \emph{optimizing for static cost metrics} problematic. Indeed, current systems all minimize predetermined costs like FLOPs or API call prices, but these commonly fail to translate to actual runtime costs (and thus, performance goals), i.e., for metrics such as throughput, resource utilization, and latency. Instead, what truly governs these aspects is how configurations fit into the current system states and utilize the currently available resources. For instance, a ``more expensive" 70B-parameter model on an idle GPU can return results faster than a ``cheaper" 14B model with a saturated request queue.

\begin{figure}[t]
    \centering
    \includegraphics[width=0.9\linewidth]{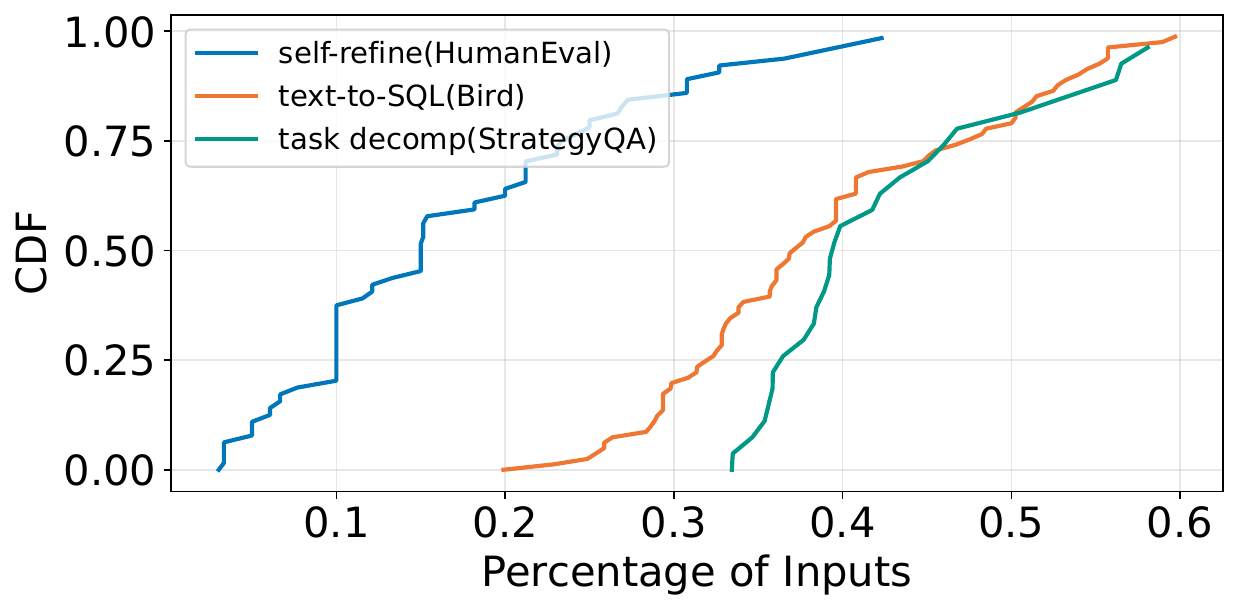}
    \vspace{3pt}
    \tightcaption{Agentic workflows exhibit robustness by recovering from intermediate errors. For each configuration, we track the inputs with intermediate errors, and record the fraction subsequently corrected by the remained workflow. The plot shows the CDF of correction percentages across all configurations.}
    \vspace{2pt}
    \label{fig:error_tolerant}
\end{figure}

\begin{figure}[t]
    \centering
    \includegraphics[width=0.9\linewidth]{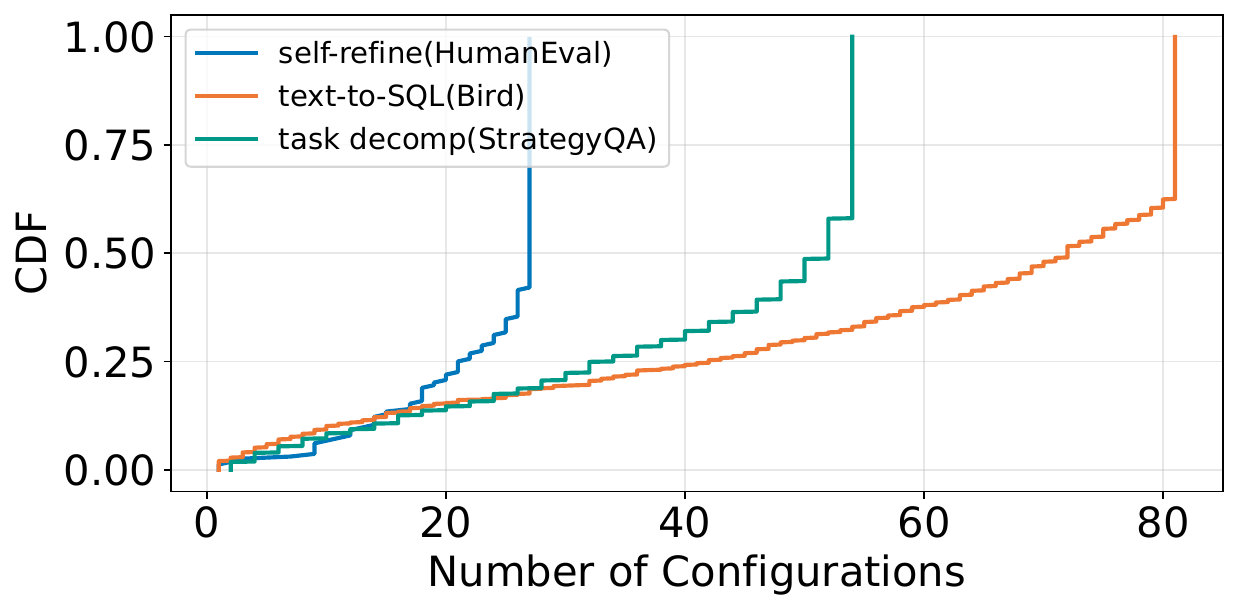}
        \vspace{3pt}
    \tightcaption{Numerous workflow configurations can match the accuracy of the most expensive configuration at for each input.}
    \label{fig:many_configurations}
\end{figure}

These drawbacks are even more problematic when contextualized relative to the unique opportunities that workflows bring. Most notably, despite the increased LLM inference overheads, the multi-stage execution of workflows provides remarkable \emph{configuration flexibility} that exceeds what is typically possible with single-model pipelines, e.g., with classic model routing. Concretely, given multiple model options per stage, agentic workflows elicit an exponentially large configuration space that presents remarkable error tolerance -- when upstream agents produce suboptimal outputs, downstream agents can often recover from those intermediate errors (Figure~\ref{fig:error_tolerant}). This self-correcting behavior dramatically expands the viable configuration space (Figure~\ref{fig:many_configurations}), highlighting an \emph{untapped potential}: the enhanced configuration flexibility provides support for frequently adapting to dynamic runtime scenarios via stage-wise configuration adaptation using the many options. By making static configuration decisions prior to workflow execution, all existing approaches fail to fully take advantage of this enhanced flexibility.

\begin{figure}[t]
    \centering
    \includegraphics[width=\linewidth]{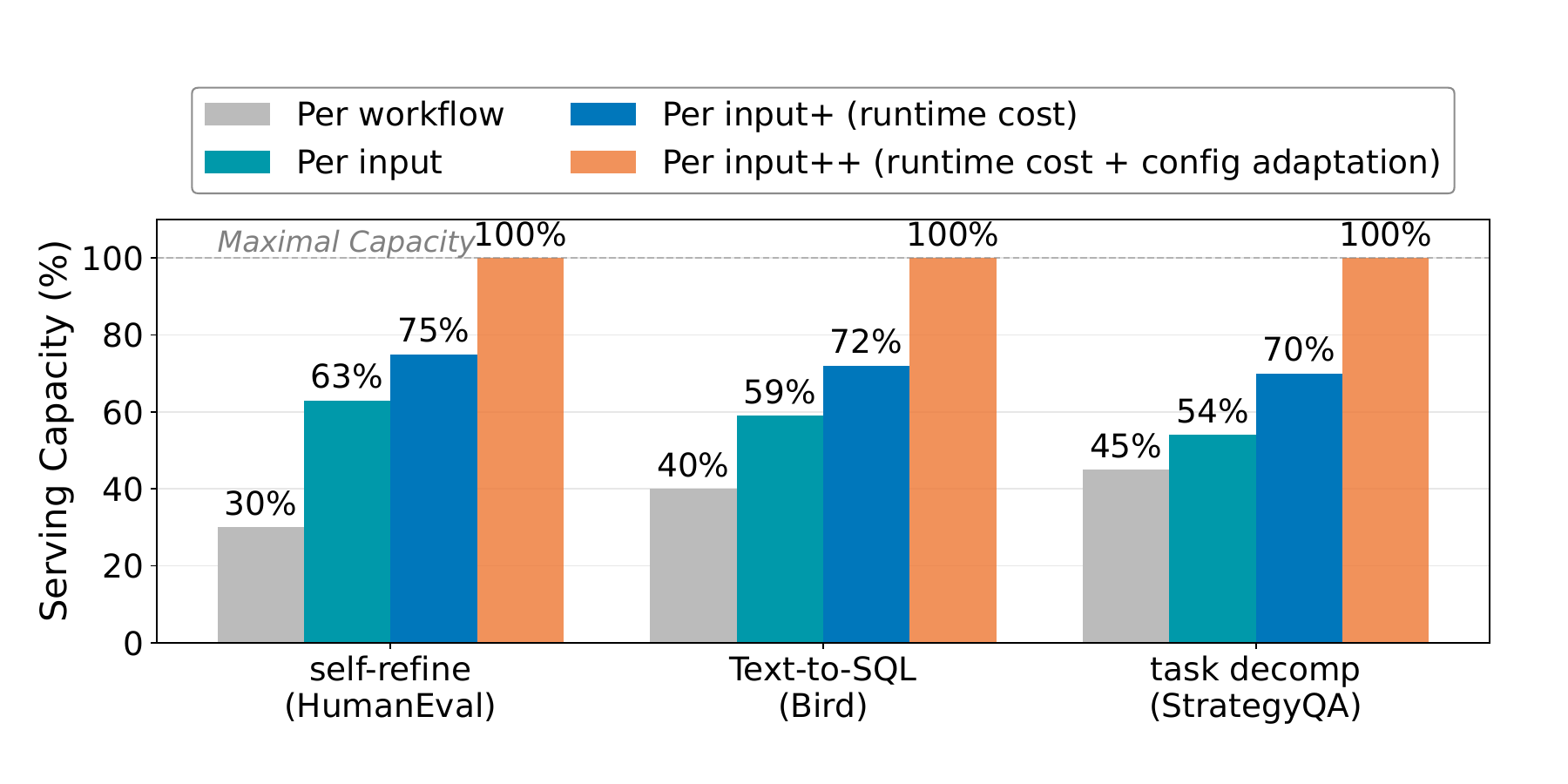}
    \caption{Static approaches (per-workflow, per-input) achieve suboptimal serving capacity. Incorporating runtime cost metrics and runtime configuration adaptation incrementally improves serving capacity. Results are evaluated on HumanEval~\cite{humaneval}, Bird~\cite{bird}, and StrategyQA~\cite{strategyQA}.}
    \label{fig:optimal_serving_capacity}
    \vspace{-15pt}
\end{figure}

To quantify this potential (and the drawbacks of existing approaches), we follow the setup from \S\ref{ss:exp_setup} and evaluate three representative workflows---self refine, text-to-sql, and task decomposition---with their corresponding datasets (HumanEval~\cite{humaneval}, Bird~\cite{bird}, StrategyQA~\cite{strategyQA}) and Qwen \{7,14,32\}B as model options. We compare the serving capacity (max achieved throughput, normalized to the best approach) of Per-workflow and Per-input configuration strategies (as described above), as well as two more schemes:
\squishlist
    \item Per-input$+$, which improves upon Per-input by factoring in runtime observations (in particular, the predicted latency based on current load for each model and serving engine) rather than solely FLOPs when selecting a configuration prior to each request, and 
    \item Per-input$++$, which improves Per-input$+$ by continuously re-evaluating and updating configuration selections at each workflow stage (rather than solely before each request) to adapt to changing system dynamics.
\squishend
To isolate impact of selection strategy, all approaches are given perfect knowledge of accuracy per configuration.

As shown in Figure~\ref{fig:optimal_serving_capacity}, despite the benefits that Per-input strategies bring via input-level configuration adaptation (i.e., 9--33\% improvement in serving capacity over Per-workflow schemes), their early binding and static cost metrics leave substantial performance gains on the table. In particular, our results highlight that incorporating dynamic cost metrics (Per-input$+$) and stage-wise configuration adaptation (Per-input$++$) can further improve serving capacity by 12--16\% and 25--30\% when they are incrementally incorporated -- a 37--46\% increase in maximum achieved throughput over even the best possible version of existing Per-input schemes.

\subsection{Challenges}
\label{ss:challenges}

Despite the benefits, realizing stage-wise configuration adaptation requires solving fundamental systems challenges in managing runtime overheads, given the large configuration space to explore and the need for frequent reconfiguration.

\para{C1: Prohibitive and frequent routing overheads.} Agentic workflows create configuration spaces that grow exponentially with workflow complexity: $M^N$ configurations for $N$ agents and $M$ models. Routing must evaluate this massive space to select configurations for each input. Yet, routing overhead scales poorly with the configuration space: state-of-the-art routers already build on large transformers for accurate single-LLM selection~\cite{routellm,mixllm}, and workflow routing demands substantially more complexity to evaluate exponentially more candidates. Moreover, adapting to rapid system dynamics requires frequent rerouting for every input. The untenable and frequent routing cost for every input can offset the benefit of flexible configuration adaptation.

\para{C2: Joint runtime reconfiguration under concurrency.} Beyond per-input routing overheads, runtime reconfiguration amplifies the coordination that serving platforms must consider when managing concurrent requests. Specifically, capitalizing on each request's configuration flexibility can constrain other requests. For instance, consider two requests where R1 can achieve target accuracy with either 7B or 14B models, while R2 needs either 7B or 32B models. If the serving engine for the 7B model can only house one more request in its batch before resorting to queuing, independent scheduling might grant R1 use of the 7B model first (its cheapest option), forcing R2 to use the expensive 32B model. In contrast, joint scheduling would recognize that R1 can make a slight performance sacrifice by using 14B, allowing R2 to take the scarce 7B slot---improving overall system throughput and latency.  However, such joint optimization becomes computationally expensive as it scales exponentially with the number of concurrent requests and their configurations.

%% file: design_v2.tex
\section{Design}
\label{s:design}

We present \name{}, an efficient agentic workflow serving system that progressively adapts each request’s configuration to cater to runtime system dynamics, i.e., resource availability. Its overarching goal is to maximize serving capacity (throughput) while preserving accuracy. To do so, our key insight is that the accuracy of a configuration for a given request remains constant, while its runtime cost (achieved throughput, latency, utilization) rapidly fluctuates with system dynamics. By recognizing this fundamental asymmetry -- accuracy (static but expensive to compute) vs. cost (dynamic but cheap to assess) -- we decouple them to make frequent runtime reconfiguration practical. Specifically, \name{} embeds (1) a \textbf{configuration predictor} (\S\ref{ss:config_prediction}) that efficiently identifies \emph{all} accurate configurations for each input \emph{once} before execution, rather than just a single configuration as in prior work, to address C1 from \S\ref{ss:challenges}, and (2) a \textbf{runtime scheduler} (\S\ref{ss:scheduling}) that progressively adapts configurations stage by stage (when actual runtime costs are observable) from only this candidate set, efficiently coordinating configuration selection across concurrent requests to maximize serving capacity (addressing C2). 

\begin{figure}[t]
    \centering
    \includegraphics[width=\linewidth]{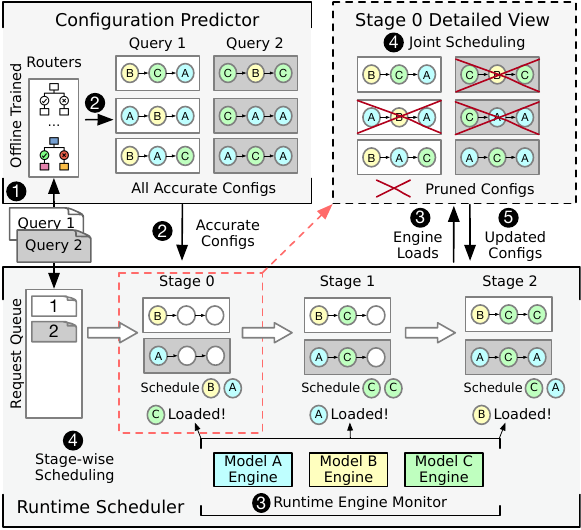}
    \caption{\name{} predicts a set of accurate configurations for each input before execution and performs stage-wise scheduling at runtime. In this stage 0 example, model C is overloaded, limiting query 2 to model A. To jointly optimize both queries, the scheduler exploits query 1's flexibility, picking configurations that start with model B.}
    \label{fig:system_overview}
    \vspace{-10pt}
\end{figure}

Figure~\ref{fig:system_overview} overviews \name{}'s architecture, which runs as a layer between agentic workflow applications and LLM serving engines. Prior to deployment, \numcircledtikz{1} \name{} trains configuration predictors using representative data for the target workflow, following existing strategies~\cite{routellm,mixllm}. Additionally, \name{} profiles the latency and throughput of serving engines under different batch sizes to estimate engine available capacity and configuration cost at runtime. Once deployed, \numcircledtikz{2} incoming requests are queued in FIFO order and first pass through the configuration predictor to obtain their set of accurate configurations. To align the stage-wise scheduling with runtime system states, the runtime scheduler periodically polls each serving engine's load \numcircledtikz{3}. When serving engines have available resources, the scheduler initiates a scheduling round \numcircledtikz{4}: using the profiled latency-throughput profiles and current load observations, it jointly selects which requests to dispatch and which models to use for their current stages such that overall serving throughput is maximized. After each scheduling round, \name{} prunes invalid configurations -- i.e., those that would have used a different model than the one selected for the current stage -- from each scheduled request's set of candidates \numcircledtikz{5}.

\subsection{Efficiently Predicting Accurate Configuration Sets}
\label{ss:config_prediction}

To effectively achieve stage-wise configuration adaptation, \name{} first needs to exploit the configuration flexibility of agentic workflows. To do that, \name{}'s configuration predictor fundamentally differs from traditional routers -- instead of selecting a single best configuration, it must identify \emph{the set} of accurate configurations for a given input. This configuration set is essential to \name{}'s operation: the prediction must be cheap enough to avoid becoming a bottleneck, while remaining accurate and presenting the maximum number of configurations to maximize adaptation flexibility.

\para{Problem formulation.} We formalize configuration prediction as follows. Given an input $x$ and a workflow DAG $G$ with $n$ agents, let $\mathcal{M} = \{m_1, ..., m_k\}$ denote the set of candidate models ordered by size (e.g., Qwen-7B through Qwen-32B), where $\mathcal{C}_\mathrm{all} = \mathcal{M}^n$ represents the full configuration space. We define $c^*$ as the most expensive configuration where all agents use the largest model $m_k$, and a configuration is \emph{accurate} if it achieves equivalent accuracy to $c^*$ for a given input. Following standard practice~\cite{routellm,mixllm,daao}, we employ learned routers to predict accurate configurations, where each router uses a two-layer architecture: an embedding model encodes the input, and is then followed by a classifier. Routers are trained offline on a labeled dataset in which, for each sample, we execute all configurations and label those matching the accuracy of $c^*$ positively.

\para{Routing strategy.} The key question is how to design routers (and the routing approach) such that accuracy is preserved but overheads are bounded, e.g., with a limit on the number of model parameters used for routing. This leads to a challenging space of options to consider. At one extreme, a single large router evaluates all $|\mathcal{M}|^n$ configurations simultaneously; at the other, $|\mathcal{M}|^n$ small binary routers each evaluate one configuration independently. Navigating this space is non-trivial: larger routers are more capable, but each must consider more configurations making its task more difficult and the accuracy-compute tradeoff unclear.

To systematically explore this space, we use three representative workflows to evaluate how overall routing accuracy scales with the number of configurations per router under a fixed total router parameter budget. For each workflow, we consider a space of 16 randomly-selected configurations of Qwen model options (7-32B) and evaluate three partitioning schemes: 16 routers each handling 1 configuration (binary routing), 4 routers each handling 4 configurations, and 1 router handling all 16 configurations, with all routers using the same train-test split. For the single-router case, we use the router described in \S\ref{ss:exp_setup}. As the number of routers increases, we scale down each router's width proportionally to maintain a constant total parameter budget.

\begin{figure}[t]
    \centering
    \includegraphics[width=\linewidth]{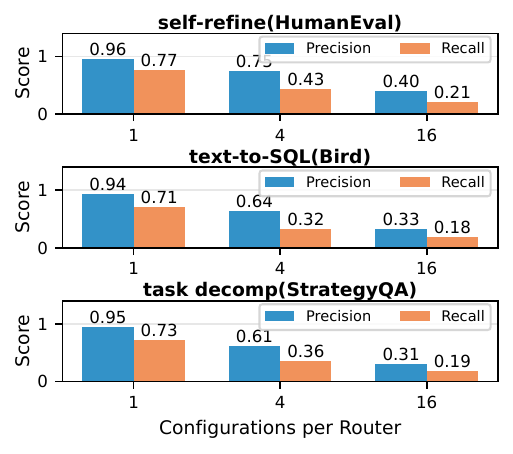}
    \caption{Overall routing accuracy drops as each router's configuration space increases. Binary routers achieve the best precision and recall for a given parameter budget.}
    \label{fig:router_scaling}
    \vspace{-10pt}
\end{figure}

Figure~\ref{fig:router_scaling} shows the overall routing accuracy results. Despite allocating the same total parameters across all approaches, the overall accuracy degrades with increasing subset size. Binary routers excel in both recall (i.e., finding more viable configurations for runtime scheduling), and even more importantly, precision (i.e., picking fewer inaccurate configurations). Specifically, binary routers achieve 94--96\% precision and 71--77\% recall, while a single router (handling 16 configurations) achieves only 31--40\% precision and 18-21\% recall. Building on this, \name{} adopts binary routers to maximize routing accuracy and focuses on enabling efficient inference of those (many) routers via two core techniques, described in turn.

\para{Pruning router inferences.} With the trained binary routers, our goal is for routing to complete during queuing for the first stage of a given request, i.e., to avoid added end-to-end serving latency from configuration adaptation. However, naively running all routers can inflate end-to-end request latency by up to 15.4\% across our workloads. 

\begin{figure}[t]
    \centering
    \includegraphics[width=\linewidth]{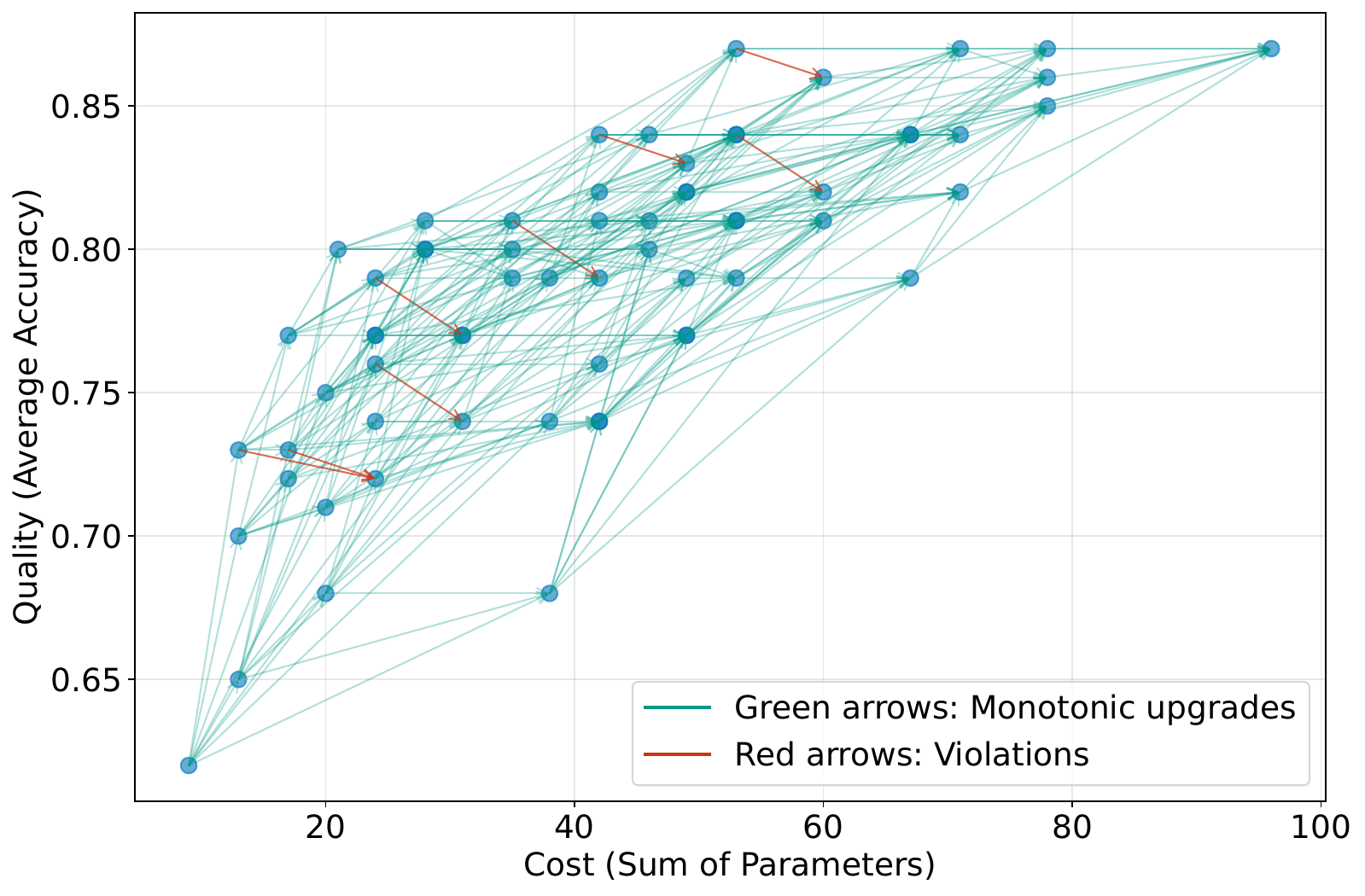}
    \caption{Workflow accuracy is largely monotonic under model upgrades. Green: valid monotonic upgrades; Red: rare violations (8 out of 288 model upgrades, with mild accuracy drops $<$ 1.5\%). The trend holds for other workflows.}
    \label{fig:monotonic}
    \vspace{-10pt}
\end{figure}

Our strategy to address this is rooted in a key empirical observation: the accuracy of a workflow exhibits near-monotonicity under model upgrades, where replacing any agent's model with a more capable one (i.e., one with more parameters) maintains or improves end-to-end accuracy in the vast majority of cases. Figure~\ref{fig:monotonic} validates this property experimentally across all configurations of a self-refine workflow used for code generation, where green arrows indicate accuracy-preserving upgrades and red arrows indicate rare violations (only 8 out of 288 upgrade cases, with accuracy drops capped at 1.5\%); note that we observe a similar trend across all other workloads in our evaluation. 

\name{} exploits this observation to carefully prune the configuration search space. Intuitively, if we run a router on a configuration and it deems a configuration accurate enough, then any configuration that upgrades one or more agents to more capable models is expected to be accurate. Conversely, if a configuration is deemed inaccurate, downgrading any agent will likely remain inaccurate. Crucially, \name{} leverages this non-guaranteed property solely to prune the set of configurations, \emph{i.e., it affects search efficiency, not accuracy}; all selected configurations still run through their corresponding routers before being considered during serving with \name{}. Thus, violations of this property can only result in reduced configuration flexibility by missing accurate configurations.

More specifically, \name{} applies this pruning strategy through a binary search. We first structure the search space into monotonic configuration chains, each starting from the base configuration where all agents use $m_1$. These chains are built via depth-first search (DFS) by individually upgrading each agent's model while keeping the others fixed. Then, along each chain, we use binary search to locate the transition boundary between inaccurate and accurate configurations. At each step in the search, we evaluate the selected configuration by running its binary router. Again, after the binary search completes, \name{} verifies all remaining configurations by running their routers and retains only those predicted to be accurate for our final candidate set.

\para{Efficient router inference.} While pruning reduces the number of routers to evaluate, executing them efficiently requires further optimization. The core issue is that the specialized classifiers from different routers cannot be batched like the shared embedding model---each classifier is router-specific, resulting in sequential execution. Worse, these classifiers are lightweight, leading to underutilized GPU resources. To address this, \name{} fuses the classifiers layer by layer via custom kernels to maximize the parallelism. 

\begin{figure}[t]
    \centering
    \includegraphics[width=\linewidth]{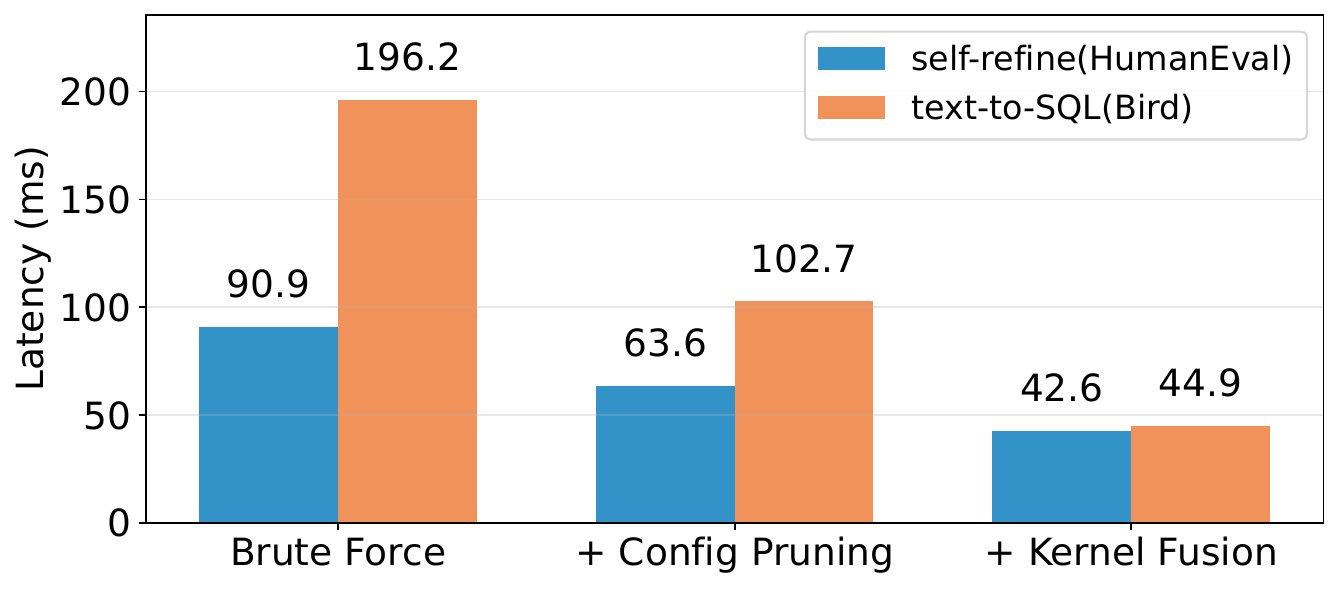}
    \caption{Reduction in average per-input routing latency from cumulative optimizations.}
    \label{fig:routing_overhead_reduction}
    \vspace{-10pt}
\end{figure}

Figure~\ref{fig:routing_overhead_reduction} shows the cumulative impact of our optimizations on average per-input routing latency for two representative workflows. Starting from naive exhaustive evaluation of $|\mathcal{M}|^n$ routers with embedding batching enabled (baseline), \name{}'s configuration pruning optimization reduces routing overheads by 30.0--47.6\%. Kernel fusion for router classifiers further reduces latency by 33.1--56.3\%. Together, these optimizations reduce average per-input routing latency from 90.9--196.2 ms to 42.6--44.9 ms per request; we show in \S\ref{ss:microbench} that the resultant overheads sufficiently ensure that routing minimally affects end-to-end response times.

\para{Overall operation.} When deployed online, \name{} adapts configuration search to the time budget -- i.e., the configuration search for a request must complete before its first stage can be executed. Specifically, \name{} estimates request queuing delays via exponential moving averages of observed per-stage queue times across all requests. Based on this estimate, \name{} adjusts the embedding model batch size and classifier fusion degree to maximize router inference efficiency while meeting the deadline. If configuration search is incomplete when a request must execute, \name{} uses the verified configurations found so far, or defaults to the most expensive configuration if none exist.

\subsection{Runtime Configuration Selection and Scheduling}
\label{ss:scheduling}

Given the set of viable configurations per request (\S\ref{ss:config_prediction}) and serving engines with provisioned resources to manage all model options, \name{}'s goal is to maximize overall serving capacity. Key to this is ensuring that stage-wise configuration decisions factor in the latest information about resource availability and also jointly optimize across all concurrent requests. 

To realize this, \name{}'s scheduler works as follows. It periodically polls each engine's current load and estimates available capacity relative to offline-profiled saturation points, i.e., the batch size for that model and serving engine beyond which throughput no longer improves. When any engine has available capacity, the scheduler triggers a scheduling round. All requests awaiting scheduling are held in a meta-queue organized in FIFO order (as in existing systems~\cite{ayo}), with each request being tagged with both its (remaining) viable configurations and its current workflow stage. With this fresh view of system state and configuration options, each scheduling round involves (1) jointly selecting requests from the queue and assigning them models to use for their current stages, and (2) dispatching those stage-wise inference jobs to the corresponding serving engines which each manage their own queues and execution logic (e.g., chunked prefill decisions). Once dispatched, requests remain in the meta-queue as pending rather than re-entering upon stage completion to preserve FIFO ordering.

\para{Joint optimization across requests and time.} \name{}'s runtime scheduler operates using the following inputs for each scheduling round: (i) $\mathcal{Q} = \{r_1, r_2, ..., r_n\}$, the request queue in FIFO order where each $r_i$ is at a specific workflow stage; (ii) $\mathcal{C}(r_i)$, the set of viable configurations for request $r_i$; and (iii) $S_m(t)$, the available slots for model $m$'s engine at time $t$ (considering current batch occupancy and maximum batch size, as per the offline profiles). In selecting request stages to schedule (and the models to use for each), \name{} strives to maximize overall serving throughput.

This involves not only maximizing current utilization across all serving engines, but also preserving future scheduling flexibility. The reason is that each scheduling round's configuration assignment has two effects: it determines resource utilization in the current round, and it potentially prunes the set of viable configurations for future stages of the considered requests. Put differently, a model selection for a given stage can render previously viable configuration options invalid (if they conflict at that stage), thereby limiting future adaptation flexibility and (potentially) compromising future resource utilization. \name{} must therefore jointly optimize across requests and time. However, the assignment space is prohibitively large -- exponential in the number of requests and the size of their configuration sets -- making exhaustive search impractical for real-time scheduling.

To handle this, our key insight is that FIFO ordering and shrinking configuration sets together narrow the assignment space for efficient exploration. First, FIFO ordering prioritizes earlier requests in the queue, and their assignments constrain the available options for later requests due to shared resources. Moreover, earlier requests in the queue are typically at later workflow stages with already-pruned configuration sets, leaving them with few viable options. Together, these two factors create a narrow search space: earlier requests have limited options, and once assigned, they further restrict later requests. This makes \emph{beam search} ideal -- it avoids both the intractability of exhaustive search and the myopia of greedy assignment by iteratively extending a small set of promising partial assignments in FIFO order.

\begin{figure}[t]
    \centering
    \includegraphics[width=\linewidth]{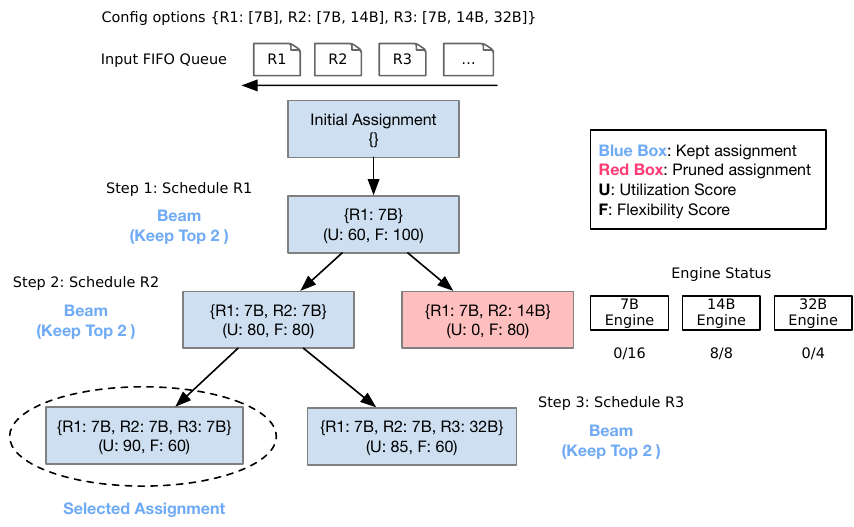}
    \caption{Beam search efficiently navigates the exponential configuration space by iteratively scheduling requests in FIFO order. Given current engine status, at each step it explores configuration options to extend the partial assignment, ranks them by utilization improvement (U) and flexibility preservation (F), and retains only the top-B (B=2 here) for further exploration.}
    \label{fig:beam_search_example}
     \vspace{-2pt}
\end{figure}

\begin{figure}[t]
    \centering
    \includegraphics[width=\linewidth]{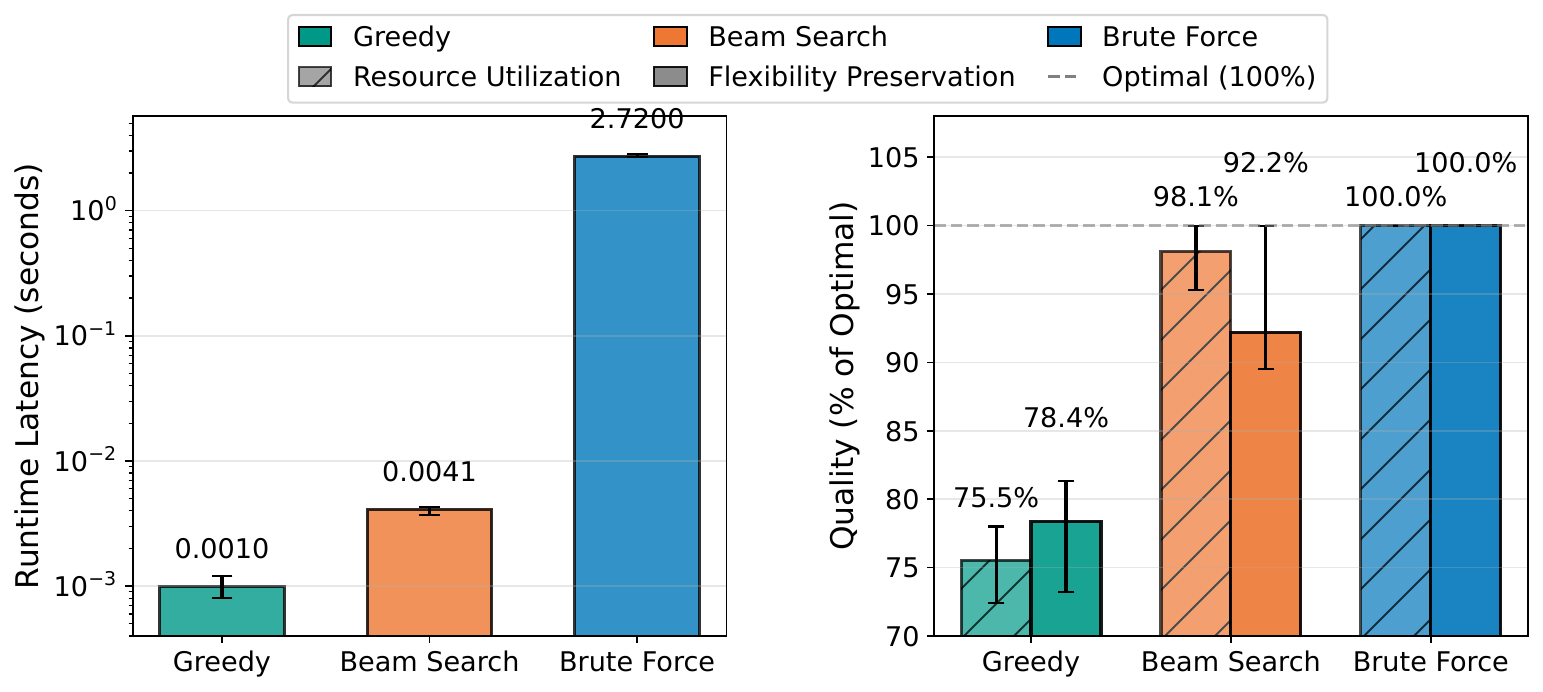}
    \caption{Comparing scheduling algorithms in \name{}: beam search achieves near-optimal resource utilization (current engine utilization) and flexibility preservation (future configuration options) comparable to brute-force, with greedy-level efficiency. Bars show averages with min-max error bars.}
    \vspace{6pt}
    \label{fig:beam_search}
\end{figure}

\name{} uses beam search to efficiently navigate the exponential configuration space (Figure \ref{fig:beam_search_example}). \name{} maintains only the top-$B$ (B is set to 4 by default, but we evaluate different values in \S\ref{ss:microbench}) most promising partial assignments at each step, where each partial assignment represents a subset of scheduled requests with their configuration assignments. Partial assignments are ranked by two criteria: \textit{(1) utilization improvement}, i.e., the total forecasted throughput improvement from the assignment, estimated using offline-profiled throughput-load curves for each model engine. This metric accounts for both each engine's inherent throughput characteristics and its current utilization. For example, the scheduler prefers assignments that distribute load across a lightly-loaded 32B engine and a 7B engine with high marginal throughput gains, rather than assignments that further saturate an already near-capacity 7B engine, and \textit{(2) flexibility preservation} as a tiebreaker, i.e., the average percentage of configuration options preserved across requests for future stages. This balances immediate resource utilization with flexibility preservation for downstream scheduling.

At each step, the scheduler expands partial assignments in FIFO order by considering all valid configuration assignments for the next unscheduled request in the queue. To avoid blocking, we employ a look-ahead mechanism (that still respects FIFO priorities): when a request cannot be scheduled due to resource constraints, the scheduler skips it and proceeds to the next request in the queue. This is critical for heterogeneous workflows where requests require different models; strict FIFO would block subsequent requests with flexible configurations behind a constrained request waiting for an overloaded model. Our relaxation preserves FIFO fairness because: (1) skipped requests add no additional delay, i.e., they remain blocked regardless of whether subsequent requests are scheduled, and (2) \name{} still iterates through requests in FIFO order in future rounds, ensuring skipped requests are never overtaken by later joint requests requiring the same models.

By retaining only top $B$ partial assignments throughout the search, \name{} makes joint optimization tractable while finding high-quality assignments. The search terminates when all feasible requests are assigned or no available serving resources remain. To evaluate the performance of beam search in \name{}, we take periodic snapshots of our evaluation workloads during serving. Figure~\ref{fig:beam_search} shows that beam search achieves near-optimal assignment quality (measured by resource utilization and flexibility preservation) that is comparable to exhaustive search while maintaining efficiency close to greedy approaches, i.e., orders of magnitude faster than brute-force search.

\para{Supporting complex workflows.} Thus far, we have assumed sequential workflows. However, agentic workflows extend beyond sequential workflows, often containing parallel branches and complex dependencies. \name{} leverages topological sorting to ensure correct execution order while maximizing parallelism. Before execution, \name{} performs a topological sort~\cite{parrot,ayo} on the workflow DAG and pre-computes each agent's depth, i.e., the longest path from that agent to any leaf node. During execution, \name{} follows a two-level priority scheme: \textit{(1) inter-request}: FIFO ordering prioritizes earlier requests over later ones, and \textit{(2) intra-request}: agents are prioritized by depth in descending order to schedule critical-path agents first. For example, given two requests $R_1$ (arrived first) and $R_2$ (arrived second), where $R_1$ has ready agents $A$ (depth=4) and $B$ (depth=3), and $R_2$ has ready agent $C$ (depth=5). Agent $A$ from $R_1$ (highest depth in earlier request) will be scheduled earlier than agent $C$, despite agent $C$ having the globally highest depth. With this ordering, \name{}'s optimization applies seamlessly to complex workflows.

%% file: implementation.tex
\section{Implementation}
\label{ss:impl}

\begin{figure}[t]
\begin{lstlisting}[language=Python, 
    basicstyle=\small\ttfamily, 
    breaklines=true,
    keywordstyle=\color{blue}\bfseries,
    commentstyle=\color{gray}\itshape,
    stringstyle=\color{red},
    emphstyle=\color{purple}\bfseries,
    emph={Workflow, ABC, dspy, Module, LM, RequestState, List, Any, int},
    morekeywords={async, await}]
class Workflow(ABC):
    @abstractmethod
    def extract_workflow_graph(
        self, dspy_program: dspy.Module
    ) :
        """Extract workflow DAG from DSPy program."""
        pass
    
    @abstractmethod
    async def execute_stage(
        self, 
        stage: int, 
        model_idx: int, 
        lm: dspy.LM, 
        request_state: RequestState
    ):
        """Execute a stage with the specified model."""
        pass
\end{lstlisting}
\caption{\name{}'s DSPy wrapper abstraction for enabling graph structure extraction and stage-wise scheduling.}
\label{fig:workflow_interface}
\vspace{-8pt}
\end{figure}

\para{}We implement \name{} as an orchestration layer in 6.8K lines of Python and Triton code that sits between agentic workflow frontends and LLM serving engines, without modifying either.

\para{Frontend and Workflow Specification.} 
As shown in Figure~\ref{fig:workflow_interface}, \name{} converts any DSPy program into its internal workflow representation, enabling runtime model reconfiguration and asynchronous stage-wise execution. We choose DSPy for its lightweight design and ease of use, making it popular for building agentic workflows. However. \name{} can support other frameworks by implementing framework-specific workflow graph extraction.

\para{Router Inference.}  We implement dynamic batching for shared embedding model serving. To accelerate router classifier inference, we develop custom Triton kernels that fuse classifiers layer by layer. We leverage CUDA graphs to further reduce the kernel launching overheads. 

\para{Runtime Scheduler.} The scheduler runs asynchronously with workflow execution. It maintains: (1) FIFO queues with look-ahead scheduling, (2) model engine states tracking occupancy, (3) configuration caches mapping requests to viable sets, and (4) exponential moving averages of queue metrics. Beam search uses priority queues, where each state encodes partial assignments as bit vectors for efficient manipulation.

\para{Serving Backend Integration.} \name{} uses SGLang~\cite{sglang}, a widely used LLM serving engine as our serving backend. However, \name{} supports any serving backends (vLLM~\cite{vllm}, TensorRT-LLM~\cite{tensorrt-llm} and HuggingFace TGI~\cite{TGI}) that provide OpenAI-compatible APIs needed for agentic workflow frontends.

%% file: eval_v1.tex
\section{Evaluation}
\label{s:eval}

We evaluated \name{} across a wide range of agentic workloads, model families, and varying request rates. Our key findings are:

\squishlist

\item \name{} improves maximum serving throughput by 42.8--76.3\% and 78.1--217.0\% over per-input and per-workflow optimizations, respectively.
\item \name{} maintains lower latency under increasing load. Under the highest load for all workloads, \name{} reduces P25, median, and P95 end-to-end latency by 23.6--60.1\%, 32.5--71.1\%, and 46.2--76.2\% over per-input optimization and 58.4--82.8\%, 60.0--86.1\%, and 63.2--89.0\% over per-workflow optimization, respectively.
\item \name{} preserves accuracy compared to always using the largest model, with maximum degradation of 2\%.
\item \name{} demonstrates robust improvements over baselines across mixed model families, router architectures, configuration space sizes, and GPU allocation schemes.
\squishend

\subsection{Experimental Setup}
\label{ss:exp_setup}

\para{Models.}
We use the Qwen-2.5 (7B, 14B, 32B)~\cite{qwen} and Llama 3 (3B, 8B, 70B)~\cite{llama3} model families. This provides a diverse range of capability-cost tradeoffs for each agent in the workflows. We use Snowflake Arctic-Embed-L-v2.0~\cite{snowflake-model} as the router's embedding model, which has a small number of parameters while providing high-quality embeddings and supporting relatively long context lengths. The router's classifier is an MLP with 6 hidden layers that maps embeddings to configuration predictions. Additionally, we explore heterogeneous model options across different families, and different embedding models in \S\ref{ss:microbench}.

\para{Testbed.}
We conduct our experiments on an H100 GPU node, which contains 8 NVIDIA H100 80GB GPUs interconnected via NVLink and a 96-core Intel Sapphire Rapids CPU. We use SGLang as our model serving engine. For models larger than 7B, we enable tensor parallelism with a degree of 2 for 14B models and 4 for 32B and larger models, following standard practice. We dedicate the remaining GPU to router inference. For baselines without routing components, this GPU instead serves an additional 7B model to ensure fair resource allocation. We also explore different initial GPU allocation configurations in \S\ref{ss:microbench}. As in prior work~\cite{vllm,sglang,autellix}, we simulate request arrivals using a Poisson process with varying arrival rates to capture different load conditions.

\para{Workflows and datasets.}
We evaluate \name{} across diverse workflow patterns (Figure~\ref{fig:workflow_examples}) and datasets. 
\squishlist
\item \textit{Self-refine for coding} on HumanEval~\cite{humaneval} (164 programming problems), using a three-agent workflow with generation, critique, and refinement agents. 
\item \textit{Self-refine for QA} on SQuAD v2.0~\cite{squad} (1000 sampled questions), following the same three-stage pattern.
\item \textit{Text-to-SQL} on Bird-SQL~\cite{bird} (780 samples), where query generation is decomposed into keyword extraction, column selection, SQL generation, and refinement stages.
\item \textit{Voting} (four voters and one aggregator) on MMLU-Pro~\cite{mmlu-pro} (1500 sampled questions), where multiple reasoning paths are generated in parallel and aggregated.
\item \textit{Task decomposition} on StrategyQA~\cite{strategyQA} (1500 sampled questions), where problems first pass through the decomposer, then invoke sub-task solvers in parallel, and finally the answer aggregator.
\squishend

\para{Baselines.}
We compare \name{} against two baselines that represent optimal performance for each category: (1) Per-workflow optimization: selects the cheapest configuration among all configurations that match the most expensive configuration's accuracy, the configuration stays fixed for all requests; and (2) Per-input optimization: performs router-based selection for each request. Both baselines are given an oracle with perfect accuracy knowledge for selecting their best configurations, representing the performance upper bound of each approach. Additionally, we remove the router inference overhead for the per-input baseline. Moreover, all baselines incorporate the same state-of-the-art graph-aware scheduling for agentic workflows as \name{} (\S\ref{ss:scheduling}).

\begin{figure*}[t]
    \centering
    \includegraphics[width=\linewidth]{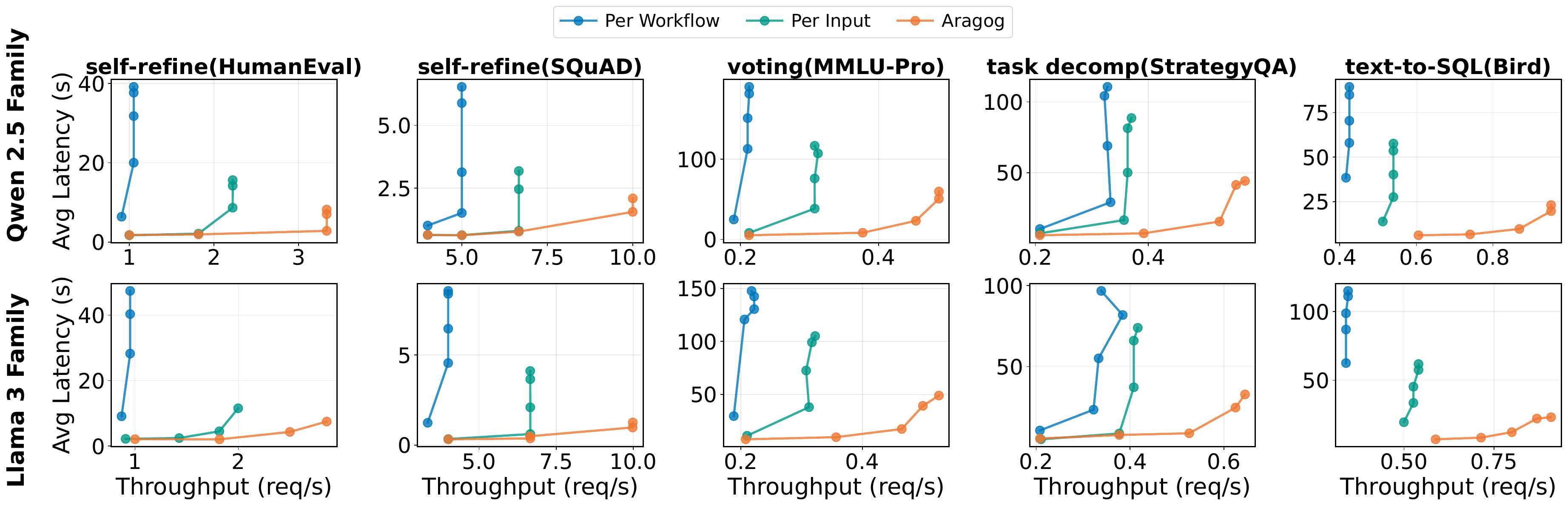}
    \caption{Average latency and throughput comparison under varying request rates and across five workflow dataset pairs and two model families. Each dot represents the average latency and throughput under a specific request rate.}
    \label{fig:serving_capacity_all}
    \vspace{-10pt}
\end{figure*}

\begin{figure*}[t]
    \centering
    \includegraphics[width=\linewidth]{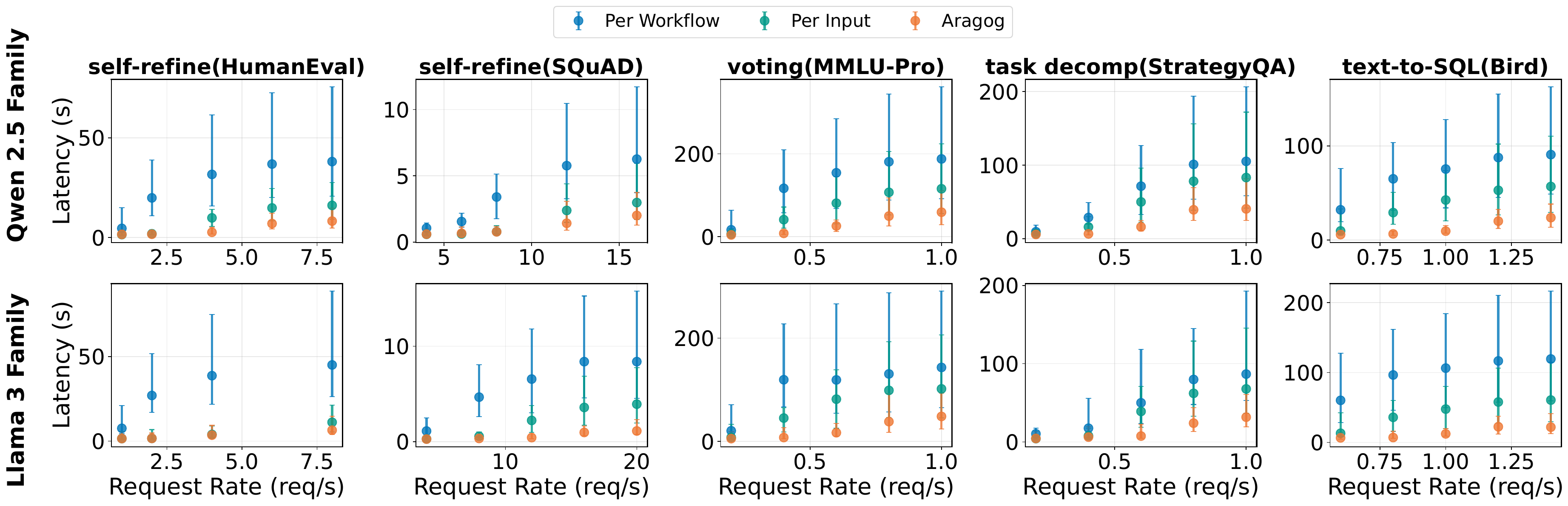}
    \caption{Median latency for workloads and model families at varying loads. Error bars show P25 and P95.}
    \label{fig:latency_reduction_all}
    \vspace{-15pt}
\end{figure*}

\para{Metrics.}
We measure serving capacity (maximal throughput), latency (Average/P25/P50/P90) under varying request rates, and accuracy preservation relative to the most expensive configuration (using the largest model for all agents).

\subsection{End-to-End Performance}
\label{ss:e2e_performance}

\begin{figure}[t]
    \centering
    \includegraphics[width=\linewidth]{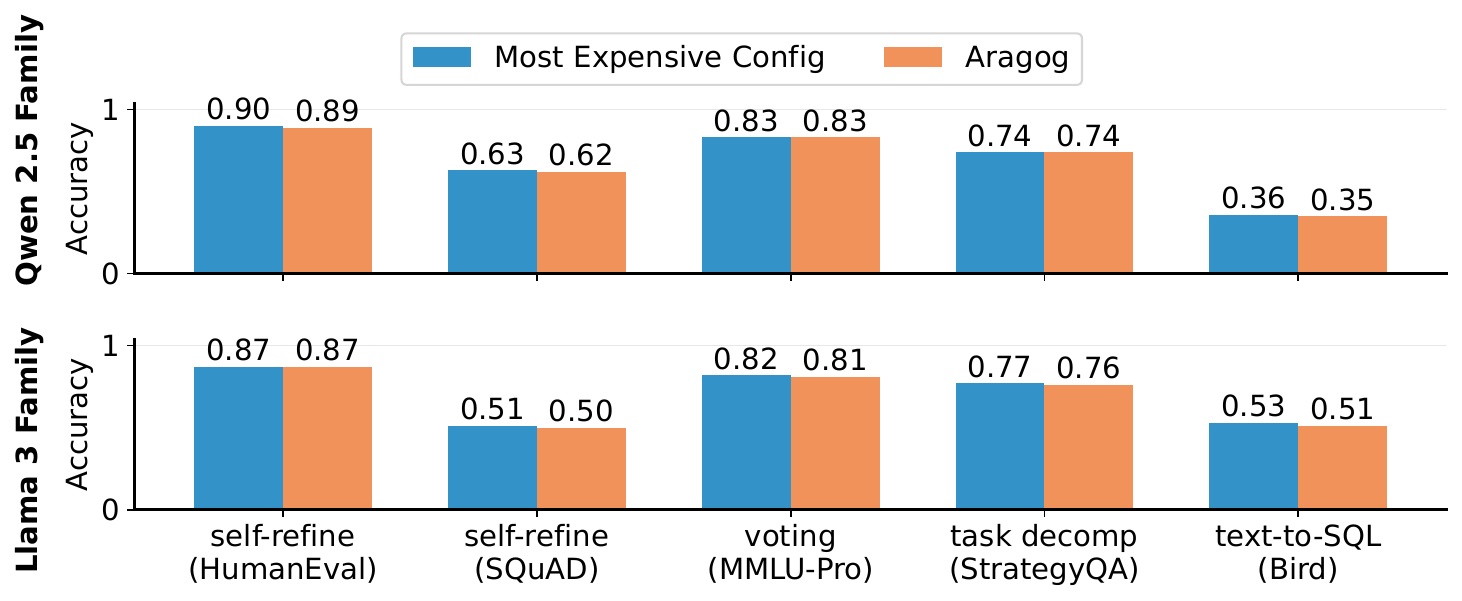}
    \caption{Evaluating \name{}'s end-to-end accuracy against the case when always using the largest model.}
    \label{fig:accuracy}
\end{figure}

Figures~\ref{fig:serving_capacity_all}--\ref{fig:accuracy} compare \name{} with two baselines across five agentic workloads and two model families. Overall, \name{} significantly improves the maximal serving capacity and lowers latencies, while matching the accuracy when always using the largest model.

\para{Serving capacity improvement.}
Figure~\ref{fig:serving_capacity_all} shows throughput-latency curves under varying request rates across workloads and model families. For each workload, we sweep the request rate from low to high and measure the achieved throughput and average latency at each rate, shown as individual points on the curves. Across both Llama and Qwen model families, \name{} consistently achieves the highest throughput and lowest average latency under varying request rates. Specifically, \name{} improves maximum serving throughput by 42.8--76.3\% over per-input optimization and 78.1--217.0\% over per-workflow optimization. Additionally, \name{} reduces average latency under the highest request rates by 25.4--69.4\% and 60.0--85.3\% over per-input and per-workflow optimization, respectively. These results show that \name{} both scales efficiently under high load and its optimization generalizes across agentic workloads and model families.

We observe several important trends across both model families. First, compared to per-input optimization, \name{}'s capacity improvements grow with workflow complexity. The throughput improvement increases from 42.8--50.1\% for self-refine workflows to 54.0--76.3\% for more complex workflows with longer sequential chains (text-to-sql) or parallel branches (task decomposition and voting). This is because workflows with more stages exacerbate the limitations of per-input optimization's static, upfront configuration. Configuration staleness increases with workflow complexity, leading to increasingly suboptimal configuration decisions. In contrast, \name{} exploits runtime reconfiguration and flexible request scheduling to maximize resource utilization across stages, yielding higher throughput gains. Second, compared to per-workflow, \name{}'s capacity improvements are more pronounced for simpler workflows (100.2--217.0\%) than complex ones (78.1--142.4\%). This is because complex workflows offer fewer opportunities for requests to exploit cheaper configurations, as tasks are harder. Similarly, the gap between per-workflow and per-input optimization narrows as workflow complexity increases.

\para{Latency reduction.}
Figure~\ref{fig:latency_reduction_all} presents latency distributions across workloads and model families under the same request rates from Figure~\ref{fig:serving_capacity_all}. At low request rates where systems are underutilized, all approaches achieve comparable latency, as expected. However, as load increases, \name{}'s latency reductions become increasingly pronounced. Under the highest load across all workloads, \name{} reduces P25, median, and P95 end-to-end latency by 23.6--60.1\%, 32.5--71.1\%, and 46.2--76.2\% over per-input optimization and 58.4--82.8\%, 60.0--86.1\%, and 63.2--89.0\% over per-workflow optimization. The improvements are most pronounced at P95, where \name{} achieves up to 89.0\% reductions. This is because through flexible runtime reconfiguration, \name{} avoids excessive queuing at individual models, reducing tail latency compared to static baselines.

\para{Accuracy preservation.}
Figure~\ref{fig:accuracy} shows that \name{} maintains accuracy across all five workloads and two model families while achieving significant throughput and latency improvements. \name{}'s accuracy matches the most expensive configuration, with differences of at most 2\%. This demonstrates that \name{}'s configuration predictor maintains high prediction accuracy while providing configuration flexibility for runtime scheduling.

\subsection{Microbenchmarks}
\label{ss:microbench}

Results here use two representative workflows: self-refine on HumanEval and voting on MMLU-Pro, using models from the Qwen family; the observed trends hold for all our workloads. Unless otherwise specified, all other experimental settings remain the same as described in \S\ref{ss:exp_setup}.

\begin{figure}[t]
    \centering
    \includegraphics[width=\linewidth]{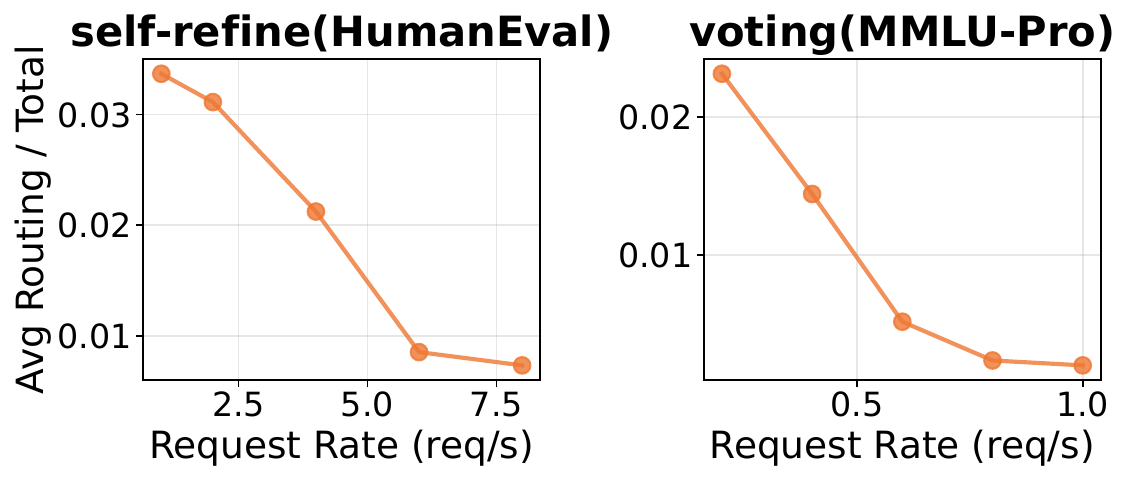}
    \caption{Ratio of average routing latency to total end-to-end latency across workloads and request rates.}
    \label{fig:system_overhead}
    \vspace{-10pt}
\end{figure}

\para{Routing overheads.} Figure~\ref{fig:system_overhead} shows the ratio of average routing latency to total end-to-end latency for different request rates. Routing overhead remains minimal, accounting for at most 3.5\% of total latency at low request rates and $<$1\% at high request rates (when queueing delays are high). However, these percentages reflect absolute overheads; with \name{}, \emph{routing does not impose any additional delay on end-to-end latencies for requests}. Indeed, at the highest request rates, routing completes 100\% of the time during queuing of each request's first stage; at the lowest request rates, \name{}'s budget-based routing (\S\ref{ss:config_prediction}) does rarely terminate routing early, but even in such cases, 90\% (on average) of routing work completes to ensure sufficient flexibility.

\begin{figure}[t]
    \centering
    \includegraphics[width=\linewidth]{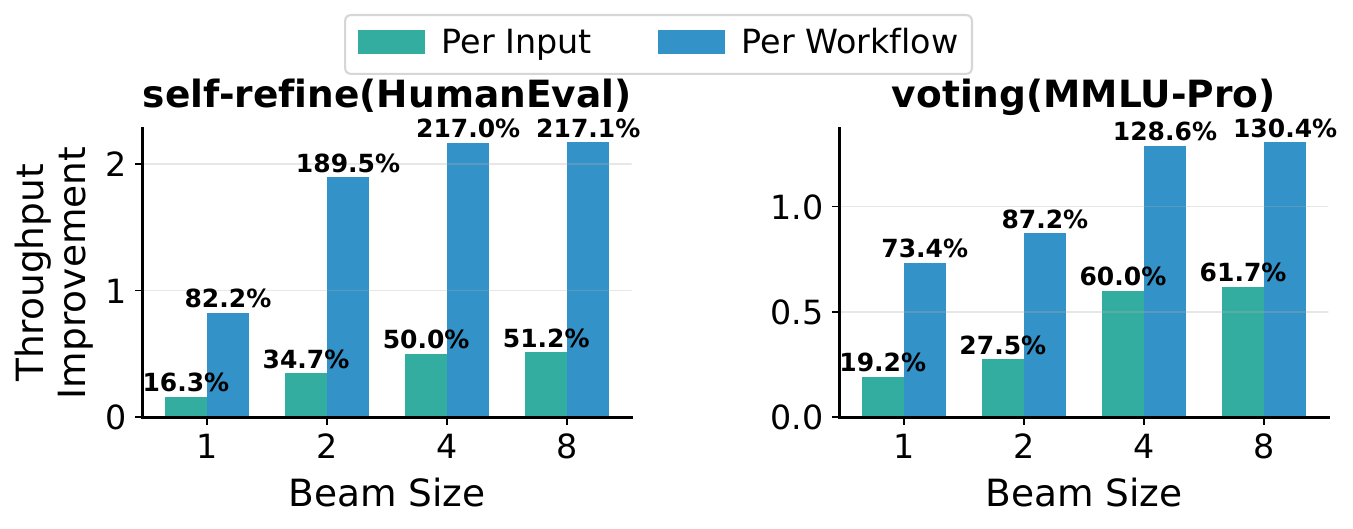}
    \caption{Impact of beam size on throughput improvement.}
    \label{fig:beam_size_impact}
    \vspace{-10pt}
\end{figure}

\para{Beam search with different beam sizes.}
Figure~\ref{fig:beam_size_impact} shows the impact of beam size on \name{}'s wins. For self-refine on HumanEval, throughput improvement over per-input optimization grows from 16.3\% to 51.2\% as beam size increases from 1 to 8, while improvements over per-workflow optimization increase from 82.2\% to 217.1\%. Similarly, for voting on MMLU-Pro, gains rise from 19.2\% to 61.7\% over per-input and 73.4\% to 130.4\% over per-workflow. Performance gains plateau with increasing beam size, as beam search finds better configuration assignments than greedy search (beam size 1) but with diminishing returns.

\begin{figure}[t]
    \centering
    \includegraphics[width=\linewidth]{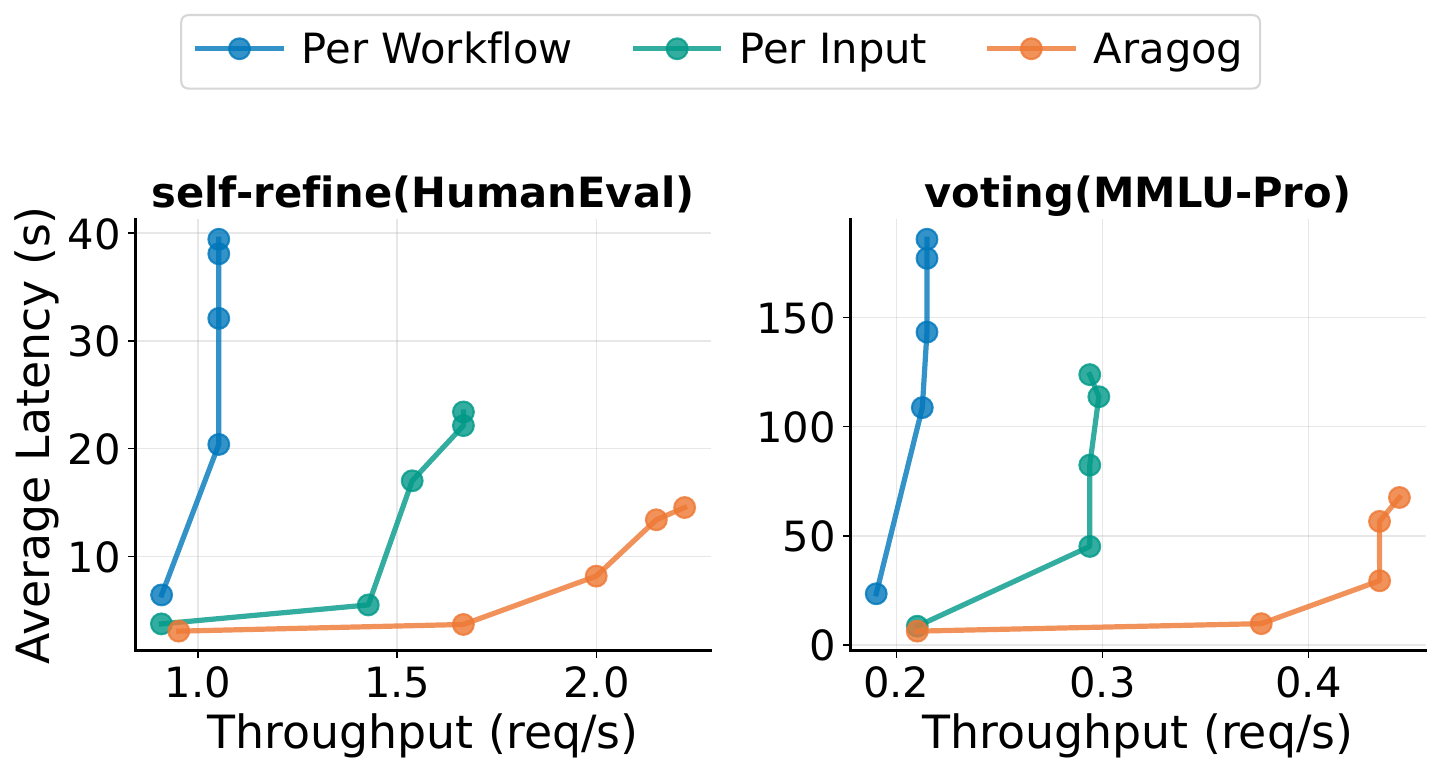}
    \caption{Average latency and throughput comparison under varying request rates using a mixed model family (Llama 8B, Phi 14B, and Qwen 32B). Each dot represents the average latency and throughput under a request rate. }
    \label{fig:mix-model-family}
    \vspace{-10pt}
\end{figure}

\para{Mixture of Model Family.}
Figure~\ref{fig:mix-model-family} shows \name{}'s generality to heterogeneous model families by using models from mixed model families: Llama3 8B, Phi4 14B, and Qwen2.5 32B. While maintaining the accuracy as the most expensive configuration, \name{} improves serving throughput by 33.7--68.5\% over per-input baseline and 109.5--111.4\% over per-workflow baseline. Additionally, \name{} reduces average latency under the highest request rates by 37.9--45.4\% and 63.1--63.6\% over per-input and per-workflow optimization, respectively.

\begin{table}[t]
\centering
\caption{Throughput improvement under different router embedding models, GPU allocations, and configuration space sizes}
\label{tab:ablation}
\small
\begin{subtable}{\linewidth}
\centering
\caption{Impact of router embedding models}
\label{tab:router_arch}
\begin{tabular}{lcccc}
\toprule
& \multicolumn{2}{c}{\textbf{Self-refine}} & \multicolumn{2}{c}{\textbf{Voting}} \\
\cmidrule(lr){2-3} \cmidrule(lr){4-5}
\textbf{Router's Embedder} & vs PI & vs PW & vs PI & vs PW \\
\midrule
Arctic-Embed-L-v2.0 & 50.0\% & 217.0\% & 60.0\% & 128.6\% \\
EmbeddingGemma & 50.3\% & 216.8\% & 58.3\% & 126.7\% \\
\bottomrule
\end{tabular}
\end{subtable}

\vspace{0.5em}
\begin{subtable}{\linewidth}
\centering
\caption{Impact of GPU allocation schemes}
\label{tab:gpu_alloc}
\begin{tabular}{lcccc}
\toprule
& \multicolumn{2}{c}{\textbf{Self-refine}} & \multicolumn{2}{c}{\textbf{Voting}} \\
\cmidrule(lr){2-3} \cmidrule(lr){4-5}
\textbf{Resource Allocation} & vs PI & vs PW & vs PI & vs PW \\
\midrule
8 H100 GPUs & 50.0\% & 217.0\% & 60.0\% & 128.6\%  \\
4 H100 GPUs & 28.3\% & 205.2\% & 32.1\% & 117.4\% \\
\bottomrule
\end{tabular}
\end{subtable}

\vspace{0.5em}
\begin{subtable}{\linewidth}
\centering
\caption{Impact of configuration space sizes}
\label{tab:config_space}
\begin{tabular}{lcccc}
\toprule
& \multicolumn{2}{c}{\textbf{Self-refine}} & \multicolumn{2}{c}{\textbf{Voting}} \\
\cmidrule(lr){2-3} \cmidrule(lr){4-5}
\textbf{Configuration Space} & vs PI & vs PW & vs PI & vs PW \\
\midrule
Full Space & 50.0\% & 217.0\% & 60.0\% & 128.6\% \\
Constrained Space & 34.2\% & 164.6\% & 52.1\% & 106.5\% \\
\bottomrule
\end{tabular}
\end{subtable}

\vspace{0.3em}
\begin{minipage}{\linewidth}
\footnotesize

All improvements measured at the highest request rate from the main evaluation (\S\ref{ss:e2e_performance}). PI = per-input baseline, PW = per-workflow baseline. All experiments maintain the same or better accuracy as the main evaluation.
\end{minipage}
\vspace{-15pt}
\end{table}

\para{Impact of router architectures.}
Table~\ref{tab:router_arch} compares two embedding models for the router: Arctic-Embed-L-v2.0 (our default) and Google's EmbeddingGemma~\cite{embeddinggemma}. Both have similar parameter counts and deliver comparable throughput performance (within 1.9\%). The negligible performance differences show that \name{}'s routing mechanism is robust to the choice of embedding model, provided that it can capture semantic similarity between workflow requests.

\para{Impact of different GPU allocation schemes.} We evaluate \name{}'s robustness to resource constraints by reducing GPU allocation from 8 to 4 H100 GPUs. With fewer GPUs, we reduce tensor parallelism accordingly: Qwen 32B from TP=4 to TP=2, Qwen 14B from TP=2 to TP=1, while Qwen 7B remains on a single GPU. This scenario favors the per-input baseline, which prioritizes smaller models based on static costs, while \name{}'s advantage comes from runtime reconfiguration that shifts load from overloaded small models to underutilized larger models. With 4 GPUs, limited capacity for larger models restricts such reconfiguration opportunities. Table~\ref{tab:gpu_alloc} shows that \name{} still achieves significant throughput improvements over the per-input baseline with 4 GPUs: 28.3\% for self-refine and 32.1\% for voting, compared to 50.0\% and 60.0\% with 8 GPUs. Gains over per-workflow optimization remain stable across both workloads (205.2\% and 117.4\% with 4 GPUs vs. 217.0\% and 128.6\% with 8 GPUs), as it selects a fixed configuration with large models to meet worst-case accuracy requirements, making it similarly resource-constrained.

\para{Impact of the size of configuration space.}
\name{} by default explores the full configuration space given a workflow and model candidates. Table~\ref{tab:config_space} investigates performance under a constrained space with additional restrictions: for self-refine workflows, the critic and refine agents must use the same model; for voting workflows, all voters must use the same model. With the constrained space, \name{} achieves 34.2\% and 52.1\% throughput improvements over per-input optimization for self-refine and voting, respectively, versus 50.0\% and 60.0\% with the full space. Gains over per-workflow optimization decrease from 217.0\% and 128.6\% to 164.6\% and 106.5\% for the two workloads. Despite reduced configuration flexibility, \name{} maintains significant advantages through joint scheduling that leverages each request's available flexibility. This also demonstrates that \name{} allows users to configure the configuration space size, which can be useful for managing complexity in larger workflows with more agents.

%% file: related.tex
\section{Additional Related Work}
\label{s:related}

\para{LLM serving optimizations.} The rapid deployment of large language models has driven significant innovations in serving optimizations. Some focus on application level compression that craft lightweight variants for efficient inference~\cite{awq,gptq,llm-pruner,speculative-decoding}. Some employ compiler techniques to optimize execution for low latency~\cite{mirage,mlc-llm}. Other works focus on system perspectives like request scheduling~\cite{orca,llmschedulinglearning}, KV cache memory management~\cite{vllm,jenga}, parallelism~\cite{alpaserve,nanoflow,magescale-infer}, efficient kernels~\cite{flashattention,flashinfer}, and resource allocation~\cite{llumnix,distserve,prism}. \name{} operates as an orchestration layer between LLM serving engines and the agentic workflow layer. As such, these optimizations are orthogonal to \name{} and can be incorporated into the underlying serving engines.

\para{Agent serving optimizations.} The rise of LLM-powered agents introduces new challenges for LLM serving, as agent execution involves an input-dependent and varying number of interceptions (e.g., tool calls) and LLM calls. Recent research has focused on addressing these agent-specific inefficiencies~\cite{autellix,infercept,pdm,tempo,asyncfunccall}. For instance, InferCept~\cite{infercept} reduces GPU resource waste arising from frequent LLM interceptions, while Autellix~\cite{autellix} mitigates head-of-line blocking caused by the variable number of LLM calls per agent by prioritizing requests based on cumulative service time. These techniques are complementary to \name{}, which orchestrates execution at the workflow level (across a graph of agents) and can leverage agent-serving optimizations within each agent within the graph.

\para{Agentic workflow serving optimizations.} Agentic workflows orchestrate multiple agents as a graph, introducing dependencies across agents. Recent systems exploit graph-level knowledge for scheduling and resource management~\cite{parrot,ayo,kairos,batchquery}. For instance, Parrot~\cite{parrot} introduces graph-aware batching and KV cache sharing, while Ayo~\cite{ayo} rearranges the execution graph for efficient execution and scheduling. These systems optimize for a fixed configuration; \name{} complements them by dynamically adapting configurations at runtime. Other work automates workflow graph design~\cite{aflow,flow}; \name{} is agnostic to graph structure and can serve automatically-generated workflows.

%% file: conclusion.tex
\section{Conclusion}
\label{s:conclusion}

We present \name{}, an efficient agentic workflow serving system that progressively adapts a request's configuration to runtime dynamics. \name{}'s core insight is to decouple accuracy and cost based configuration selection. The expensive accuracy-based configuration prediction happens once and before execution, while the lightweight cost-based configuration selection runs frequently throughout execution. \name{} makes runtime reconfiguration efficient with novel acceleration strategies. Across diverse agentic workflows and model families, \name{} achieves 42.8–217.0\% higher serving capacity than baselines while maintaining the accuracy of the most expensive configuration. \name{} delivers lower serving latency under different request loads. Under peak load, it delivers 32.5--86.1\% and 46.8--89.0\% lower median and P95 latency, respectively.